\documentclass[fleqn,twoside]{article}
\usepackage{espcrc2}

\usepackage{graphicx}
\usepackage[figuresright]{rotating}

\newcommand{\AmS}{{\protect\the\textfont2
  A\kern-.1667em\lower.5ex\hbox{M}\kern-.125emS}}

\title{
Dynamical model of electroweak pion production in the resonance region
}

\author{
 T. Sato\address{
Department of Physics, Osaka University, Toyonaka, Osaka 560-0043,
 Japan},
 B. Szczerbinska\address[USC]{
Department of Physics and Astronomy, University of South Carolina,
Columbia, SC 29208, USA},
 K. Kubodera\addressmark[USC]
 and
T. -S. H. Lee\address{
Physics Division, Argonne National Laboratory, Argonne, IL  , USA.}
}
       
\begin{document}

\begin{abstract}
In this report, we will briefly review the dynamical model of pion electroweak 
production reactions in the $\Delta$ resonance region and report on our 
study of neutrino-nucleus reactions based on this model.

\vspace{1pc}
\end{abstract}

\maketitle

\section{Introduction}

It is well recognized that the precise knowledge of neutrino-nucleus 
reactions is crucial in analyzing neutrino oscillation experiments~\cite{sakuda}.
For neutrino  reactions around 1 GeV, quasi-elastic 
scattering and pion production processes are the main reaction mechanisms.
The pion production takes place through both non-resonant and resonance processes, 
and in the latter the $\Delta_{33}$ resonance plays a central role 
for neutrinos in the 1 GeV region.
Since experimental data on the neutrino-nucleon 
reactions in the resonance region
are not as extensive as electron scattering data, a
theoretical study 
of the weak pion production amplitude will be valuable for the
study of neutrino-nucleus 
reactions~\cite{sl1,sl2,sl3,sl4,pascos2,pascos1}.

In the recent years,
extensive  studies 
on the electron- and photon-induced meson production reactions
in the GeV region are in progress to investigate 
the nucleon resonance properties~\cite{lb}. 
The objective of the $N^*$ study is to understand the
non-perturbative dynamics  of QCD by testing the resonance properties 
against the prediction of QCD-inspired models
and/or lattice simulations.
Two of the present authors with their collaborators
have developed a dynamical model for
describing photo- and electro-production of pions off the 
nucleon  around the $\Delta$ resonance region~\cite{sl1,sl2}.
It has been shown that
including the pion cloud effect in theoretical analyses
can resolve a long-standing puzzle that 
the $N\Delta$ magnetic dipole
transition form factor $G_M$ predicted by the quark model is
about 40\% lower than the empirical value.
Furthermore, the predicted  E2 $G_E$ and C2 $G_C$ form factors
show pronounced $Q^2$ dependence due to the pion cloud effects, 
which suggests the deformation effects in the $N\Delta$ transition.
This model is further extended to investigate 
neutrino reactions, which involve the axial-vector responses
of hadrons~\cite{sl3,sl4}.
The $N\Delta$ axial vector form factor $G_A$ is found to contain
large meson cloud effects.

The dynamical model starts from the non-resonant meson-baryon
interaction and the resonance interaction, and the 
unitary amplitudes are obtained from the scattering equation.
Fairly consistent descriptions of all the available data in the
$\Delta$ resonance region have been obtained from the dynamical
model.  The well tested dynamical model 
for the single-nucleon case
will be a good starting point for investigating 
the reaction dynamics of neutrino-nucleus reactions.

In this paper we give a short review of our dynamical model 
for meson production reactions in the delta resonance region.  
Subsequently we show the results of our first application of this model
to neutrino-nucleus reactions 
in the delta production region.

\section{Dynamical model}

We briefly describe our dynamical model of pion photoproduction.
We start from the interaction Hamiltonian ($H$) for 
the $N, \Delta, \pi, \rho$ and $\omega$ fields. By assuming 
that only `few-body' states are active in the
energy region we are interested in, 
we can simplify the intrinsic many-body problem.
The effects due to many-body states are
absorbed in effective interaction operators. 
In our work, the effective `few-body' Hamiltonian  $H_{eff}$ 
is derived from $H$ using the unitary transformation method~\cite{koba}.
The resulting effective Hamiltonian $H_{eff}$ is defined in a subspace
spanned by  $\pi N$, $\Delta$, $\gamma N$ states and has the following form:
\begin{eqnarray}
H_{eff}  =  H_0 + 
\Gamma_{\pi N \leftrightarrow \Delta} + \Gamma_{\gamma N \leftrightarrow \Delta}
 + v_{\pi N} + v_{\gamma N}.
\end{eqnarray}
Here $v_{\pi N}$ and $v_{\gamma N}$ are non-resonant interactions,
which consist of 
the u- and s-channel nucleon exchange and t-channel
pion and vector meson exchange processes shown in Fig.\ 1(a).
The excitation of $\Delta$ is described by 
$\Gamma_{\pi N \leftrightarrow \Delta} $
and $\Gamma_{\gamma N \leftrightarrow \Delta}$ shown in Fig.\ 1(b).
An important feature of our approach is that the effective
Hamiltonian is energy independent. Hence the unitarity of the
resulting amplitude is automatically satisfied.  Furthermore, the
non-resonant interactions $v_{\pi N}$ and $v_{\gamma N}$ are
derived from the same unitary transformation 
and hence 
$\pi N$ and $\gamma N$ reactions can be described consistently.

\begin{figure}[htb]
\includegraphics[width=7cm]{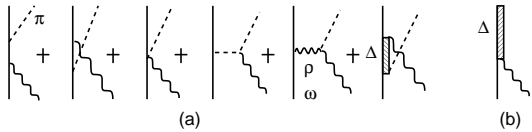}
\label{fig-h}
\caption{Graphical representation of $\gamma N \rightarrow \Delta$ and
 $\gamma N \rightarrow \pi N$ interactions.}
\end{figure}
From the effective Hamiltonian, it is straightforward to derive
a set of coupled equations for $\pi N$ and $\gamma N$ reactions.
The resulting pion photoproduction amplitude is given as
\begin{eqnarray}
  T_{\gamma \pi}& = & <\!\pi N|\epsilon\cdot J^{em}|N\!> \nonumber \\
 & = & t_{\gamma \pi}(W) + 
\frac{\bar{\Gamma}_{\Delta \rightarrow \pi N}(W){\bar{\Gamma}_{\gamma N
\rightarrow \Delta}}(W)}{W - m_\Delta^0 - \Sigma(W)}, \label{heff}
\end{eqnarray}
where $W$ is the invariant mass of the $\pi N$ system.
The first term is the non-resonant amplitude $t_{\gamma\pi}$
calculated from the
non-resonant interactions $v_{\pi N}$ and $v_{\gamma N}$.
The second term in Eq. (\ref{heff}) is the resonant amplitude.
The self-energy $\Sigma$ is defined as
\begin{eqnarray}
\Sigma(W) = \Gamma_{\pi N \rightarrow \Delta} G_0 \bar{\Gamma}_{\Delta
 \rightarrow \pi N}(W).
\end{eqnarray}
The `dressed'  $\gamma N \rightarrow \Delta$ vertex is defined by
\begin{eqnarray}
\bar{\Gamma}_{\gamma  N \rightarrow \Delta}(W)
 = \Gamma_{\gamma  N \rightarrow \Delta} + 
  \int \Gamma_{\pi  N \rightarrow \Delta} G_0 t_{\gamma \pi }(W).
\end{eqnarray}
An important consequence of the dynamical model 
is that the influence of
the non-resonant mechanisms on the resonance properties can be
identified and calculated explicitly.
The `bare' resonance vertex $\Gamma_{\gamma N \rightarrow \Delta}$ is
modified by the non-resonant meson cloud $t_{\gamma \pi}$ to give
the dressed vertex $\bar{\Gamma}_{\gamma N \rightarrow \Delta}$.

\section{Pion electroweak production and $N\Delta$ form factors}

The dynamical approach described in the previous section
was used to study $\pi$-$N$ scattering,
pion photoproduction and pion electroproduction
reactions~\cite{sl1,sl2}.
The model was subsequently extended to investigate neutrino
induced reactions~\cite{sl3,sl4}.  
Since the weak currents in the standard model
are closely related to the electromagnetic current, 
it is straightforward to extend the
dynamical model of pion electroproduction to 
weak pion production reactions.
The electromagnetic current ($j^{EM}_\mu$), 
the weak charged current ($j^{CC}_\mu$) and 
the weak neutral current ($j^{NC}_\mu$)  can be expressed 
in terms of the vector current $V_\mu$ and the axial-vector current $A_\mu$ 
as
\begin{eqnarray}
  j^{EM}_\mu & = & V_\mu^3 + V_\mu^{Iso-Scalar}, \\
  j^{CC}_\mu & = & V_\mu^{1+i2} - A_\mu^{1+i2}, \\
 j^{NC}_\mu & = &  (1-2\sin^2 \theta_W)j_\mu^{em}  
            - V_\mu^{IS} - A_\mu^3.
\end{eqnarray}
Using CVC, $V_\mu^{1,2}$
can be obtained from $V_\mu^3$ by isospin rotation.
Guided by the effective chiral Lagrangian method and
using the unitary transformation method, we can construct
$A_\mu$ for pion production.  
The resulting $A_\mu$ consists of
the nucleon-Born term, rho-exchange and delta excitation terms. 
The `bare' $N\Delta$ magnetic and electric form factors $G_M(0)$
and $G_E(0)$ are obtained from the analysis of 
the $(\gamma,\pi)$ reaction.  
We assume the Siegert theorem to determine the 
Coulomb quadrupole form factor $G_C(0)$ from 
the transverse electric form factor $G_E(0)$. 
The $Q^2$ dependence of the
form factors are assumed to be
\begin{eqnarray}
G_\alpha(Q^2) = G_\alpha(0) R_{SL}(Q^2)G_D(Q^2), \label{ff}
\end{eqnarray}
where  $G_D = 1/(1 + Q^2/m_V^2)^2$ is 
the dipole form factor of the proton with $m_V^2=0.71\,{\rm GeV}^2$.
The correction factor $R_{SL}$ is defined as
\begin{eqnarray}
R_{SL}(Q^2) & =& (1 + a Q^2) \exp(- b Q^2),
\end{eqnarray}
where $a=0.154\,{\rm GeV}^{-2}$ and $b=0.166\,{\rm GeV}^{-2}$ 
have been deduced by fitting the
$(e,e'\pi)$ data at $Q^2=2.8$ and $4\,{\rm GeV}^2$~\cite{fro}.
For  the $N\Delta$ axial vector form factor  $G_A(Q^2)$,
we also assume the form given in Eq. (\ref{ff})
with $G_D(Q^2)$ replaced by the nucleon 
axial-vector dipole form factor  
with $m_A=1.02$ GeV.
We further assume  
the SU(6) quark model relation 
between $G_A$ for the $N\Delta$ transition and $g_A$ 
of the nucleon. Having determined 
the axial $N\Delta$ form factor,
we have no further adjustable parameter in our model
to describe neutrino-induced pion production reactions.

In Fig.\ 2,  \,\,the energy dependences of the
$p(e,e'\pi^+)n$ and $p(\nu,e^-\pi^+)p$
cross sections are shown at $Q^2=0.1({\rm GeV}/c)^2$ and $E_{lepton}=1$ GeV. 
The solid lines show the full results and the
dashed lines show the contribution of the resonance
amplitude which is the second term of Eq. 2.
Clearly the non-resonant amplitude plays
a crucial role. It is well known that 
the Kroll-Ruderman term dominates 
low-energy charged-pion production reactions.
We note that even in the $p(\nu,e^-\pi^+)p$ reaction, where
only the isospin 3/2 amplitude contributes, 
the non-resonant contribution to the cross section
amounts to  15\% at $W=1.2$ GeV.
Thus the appropriate treatment of the 
non-resonant mechanisms  plays 
an important role in interpreting the $N\Delta$ transition form factors 
$G_\alpha(Q^2)$ and also 
in determining the neutrino-nucleon reaction amplitudes
that are needed for calculating the neutrino-nucleus
reaction amplitudes.

\begin{figure}[hbt]
\hspace*{-0.1cm}\includegraphics[width=4cm]{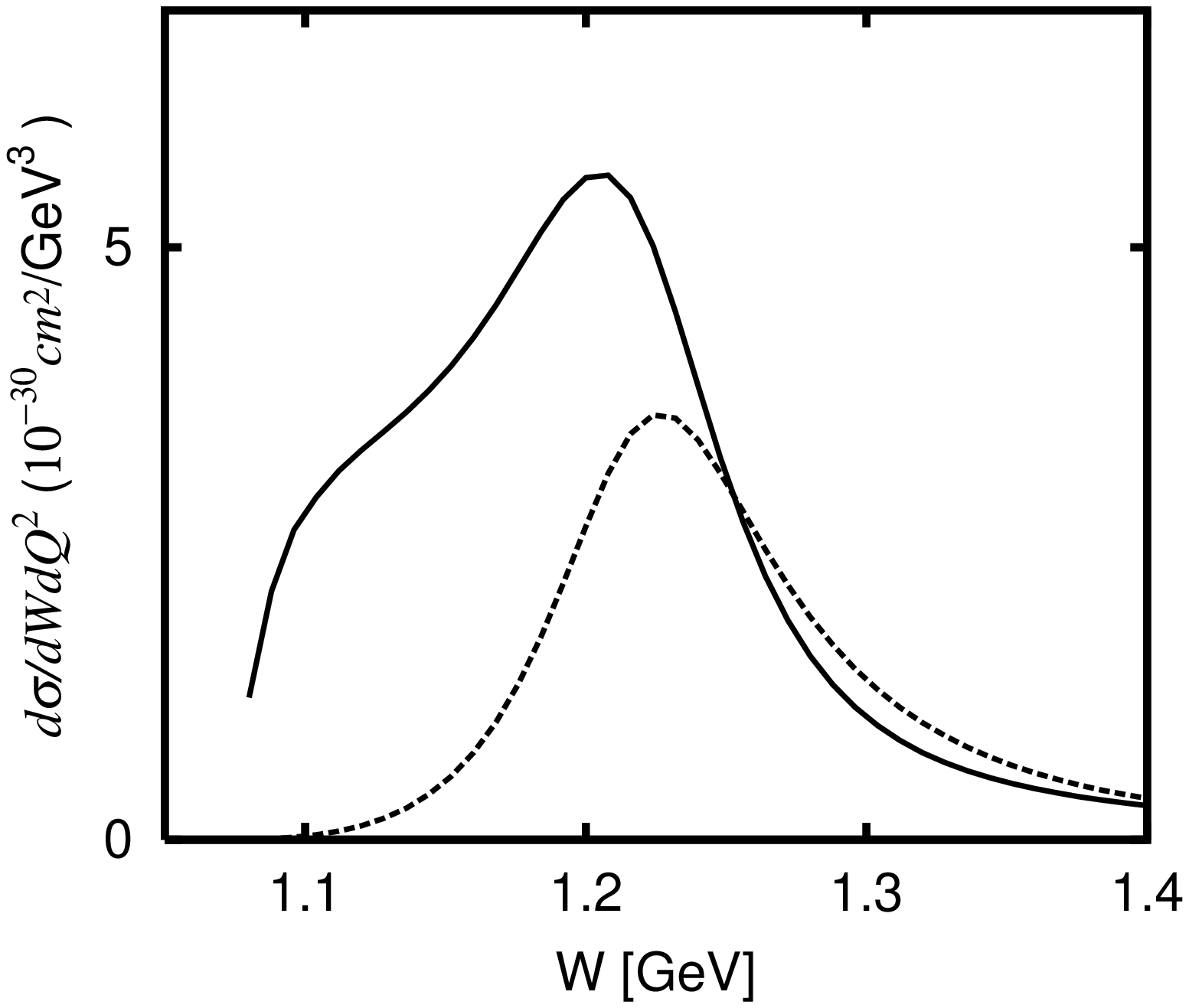}\hspace*{-0.1cm}
\includegraphics[width=4cm]{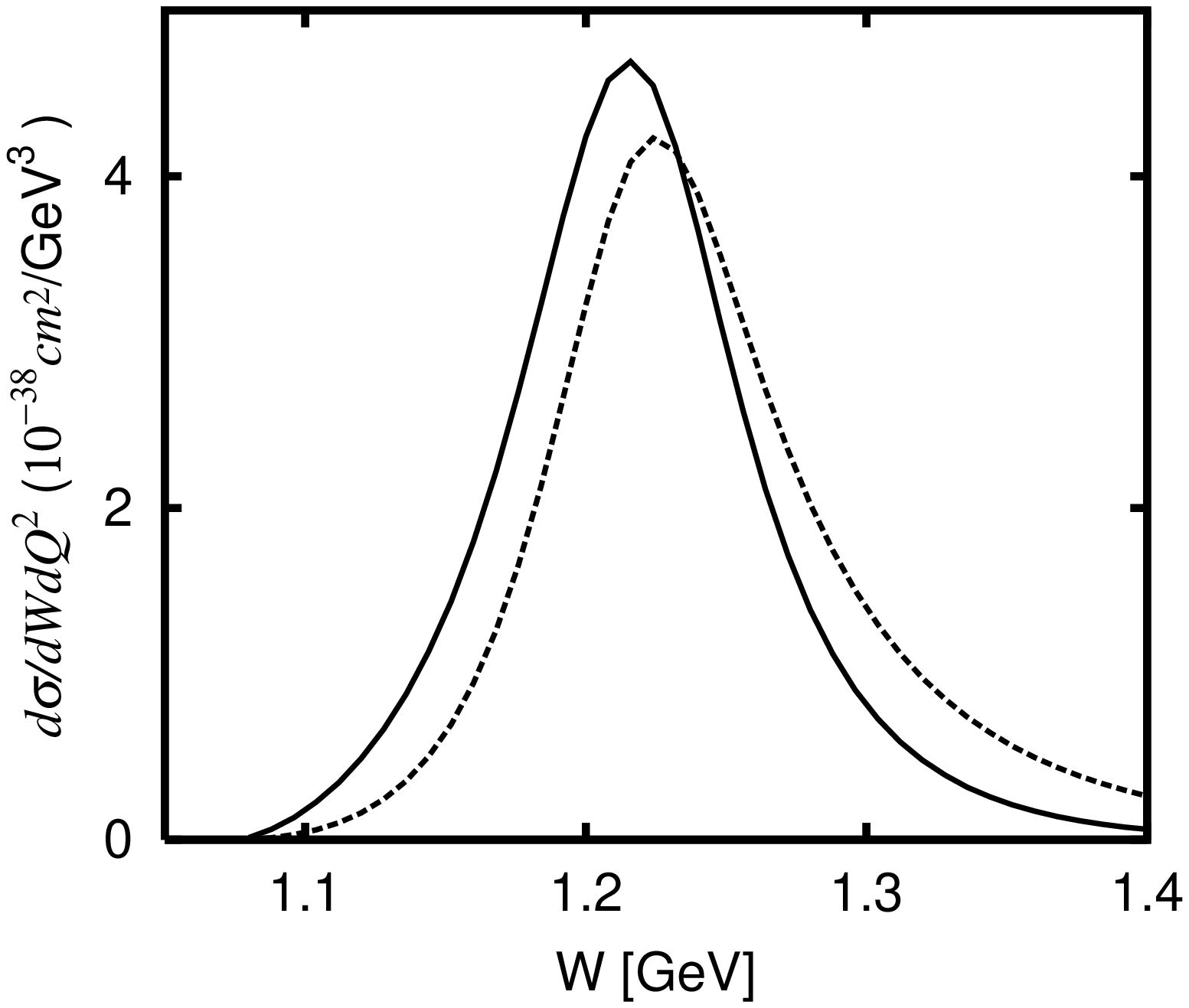}
\caption{Cross section $d\sigma/dW/dQ^2$ of 
$p(e,e'\pi^+)n$ (left) and $p(\nu,e^-\pi^+)p$ (right)
reactions. }
\end{figure}

The pion production mechanism of 
our dynamical model may be
tested by comparing the theoretical and experimental values
of the cross sections for a wide energy ranges also useful.
For instance, the longitudinal-transverse interference term 
$\sigma_{LT'}$ for a polarized electron is given by 
interference between the real and imaginary parts
of the amplitude
and hence it is sensitive to the non-resonant mechanism.
As an example, we compare in Fig. 3
our predictions for the
$e + p \rightarrow e' + p + \pi^0$ reaction
with the Jlab data~\cite{joo}.
The cross section for virtual-photon pion production
can be expressed as
\begin{eqnarray}
\frac{d\sigma}{d\Omega_\pi } &=&
[\frac{d\sigma_T}{d\Omega_\pi } + \epsilon \frac{d\sigma_L}{d\Omega_\pi
} ]
+ \sqrt{2\epsilon(1+\epsilon)}\frac{d\sigma_{LT}}{d\Omega_\pi }\cos
\phi_\pi \nonumber \\
&& + \epsilon \frac{d\sigma_{TT}}{d\Omega_\pi } \cos 2\phi_\pi.
\end{eqnarray}
The transverse
cross section $\sigma_T$ is mainly determined by
the contribution of the magnetic dipole form 
factor $G_M$.
Interference between the transverse and longitudinal
currents
($\sigma_{LT}$) is sensitive to the product $G_C G_M$.
Our model gives a reasonable description of the angular 
distribution from threshold to the delta resonance region.
Reasonable agreement has also been seen
in the comparison of our dynamical model predictions
with the extensive pion production data in the delta resonance 
region obtained at LEGS, Mainz, Jlab and MIT-Bates.

\begin{figure}[htb]
\includegraphics[width=8cm]{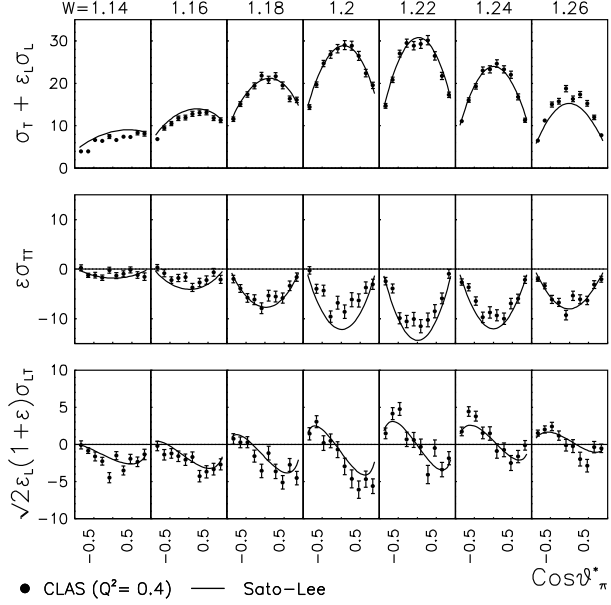}
\label{fig-eepi}
\caption{$p(e,e'\pi^0)p$ cross section at $Q^2=0.4 ({\rm GeV}/c)^2$.}
\end{figure}

We compare in Fig. 4 the calculated
$Q^2$ dependence of the cross section
for the
$\nu_\mu + p \rightarrow \mu^- + \pi^+ + p$  reaction
with  the ANL~\cite{anl} and BNL~\cite{bnl} data. 
Here we take into account 
the variation of the neutrino flux 
in the  experiment and the finite mass of 
the muon. The results of our full calculation (solid line) agree
reasonably well with the
data both in magnitude and $Q^2$ dependence.
The contribution of 
the axial vector current (dashed curve)
and the vector current (dot-dashed curve) 
have rather different
$Q^2$ dependences in the low $Q^2$ region and 
interfere constructively
with each other. 
Since the vector current contributions are 
highly constrained by the $(e,e'\pi)$ data, the results
in Fig. 4 suggest that our model for
the axial vector current reaction
is consistent with the data.
This in turn implies that the axial-vector $N\Delta$ coupling 
constant predicted by the SU(6) quark model 
is consistent with the existing data of neutrino-induced
pion production reactions in the $\Delta$ resonance region. 
A precise measurement of parity-violating asymmetry of inclusive 
$p(\vec{e},e')$ can be used to improve the determination of the
axial $N\Delta$ form factor~\cite{sl4}.

\begin{figure}[hbt]
\includegraphics[width=3cm]{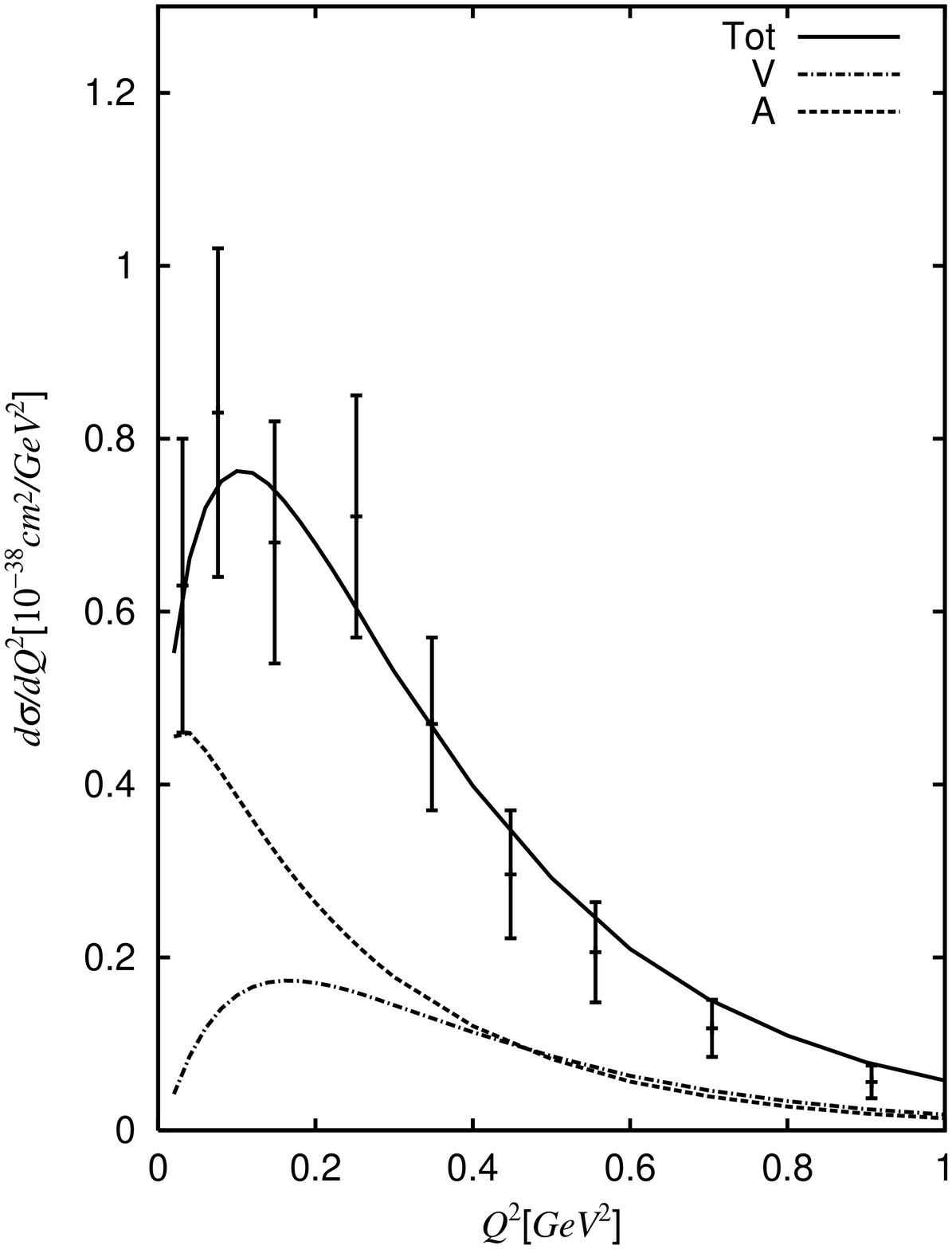}\hspace*{-0.1cm}
\includegraphics[width=4.2cm]{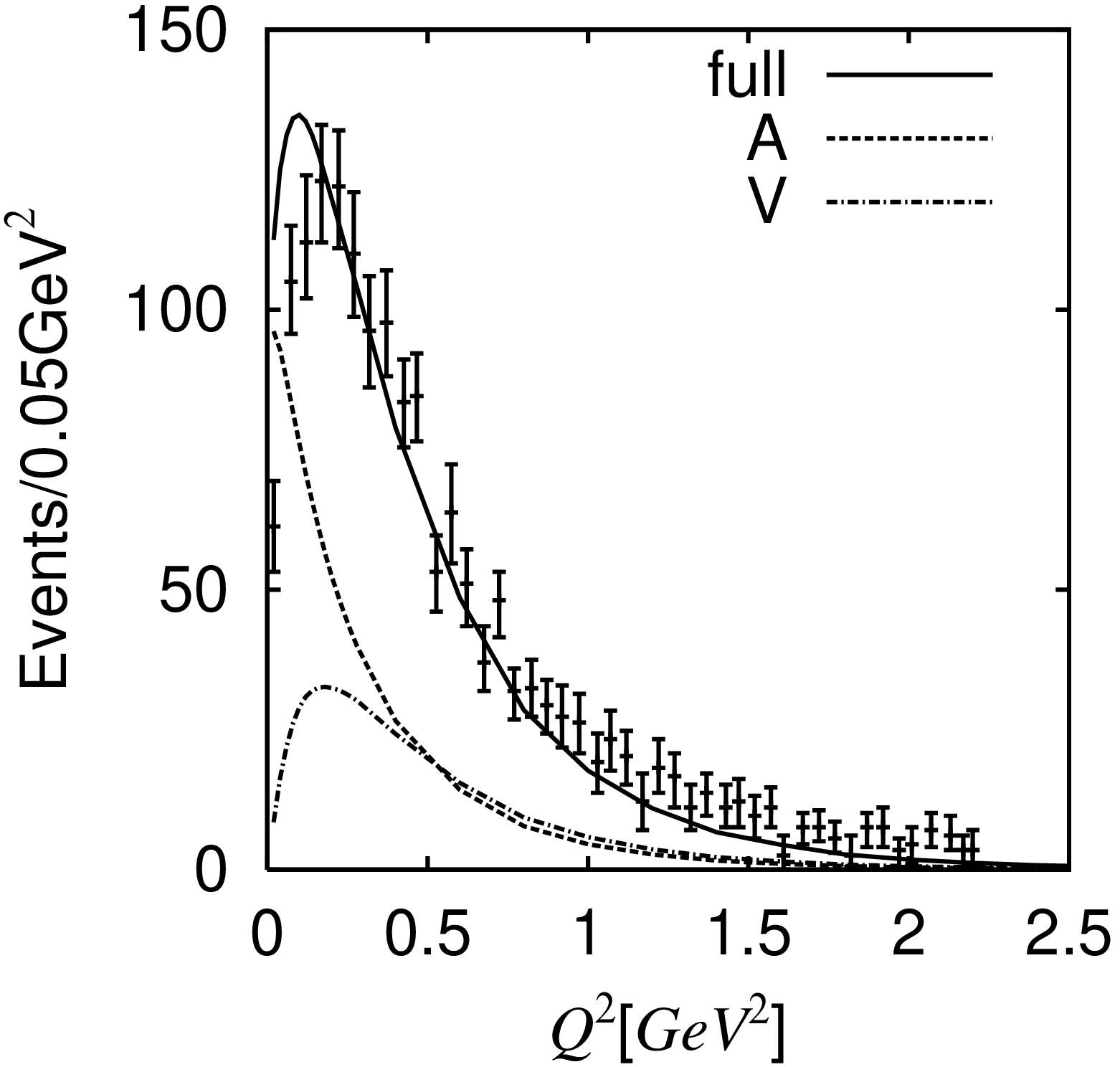}
\caption{Differential cross section $d\sigma/dQ^2$ of $p +\nu_\mu
\rightarrow \mu^- + \pi^+ + p$ reaction}
\end{figure}

Finally, we explore the effects 
of 
the pion cloud
on the $N\Delta$ transition form factors.
Fig.\ 5 displays
the dressed magnetic dipole form factor, $G_M$, and
the dressed axial-vector form factor, $G_A$,  extracted from 
our present model.
The pion cloud effect is 
seen to be
essential in explaining the
empirical values of $G_M$ directly extracted from the data.
The pion cloud effects are
sizable also for $G_A$. 
The dynamical model 
explains
why most of the quark 
models underestimate
$G_M$ and $G_A$.

\begin{figure}[hbt]
\includegraphics[width=4cm]{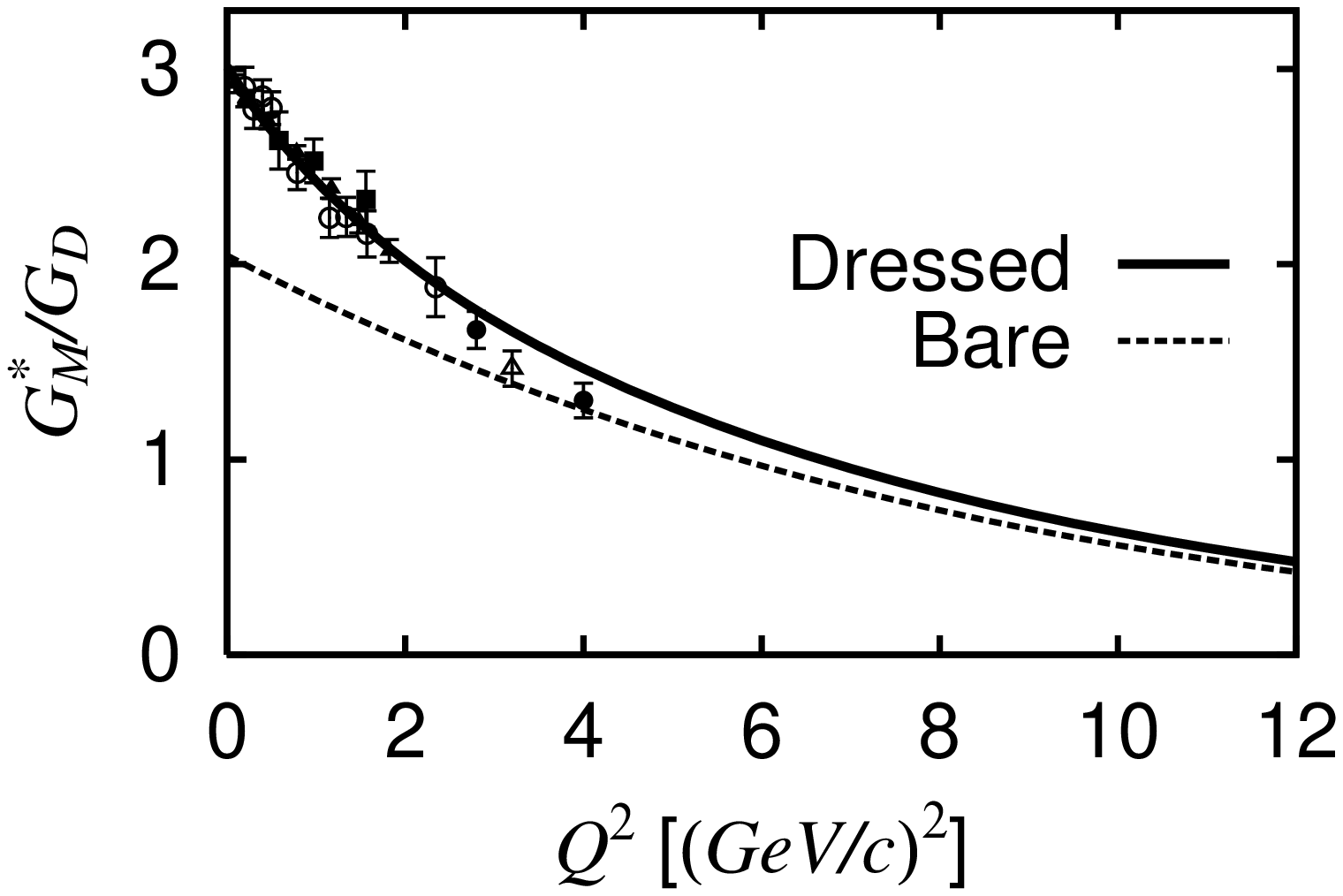}
\includegraphics[width=3cm]{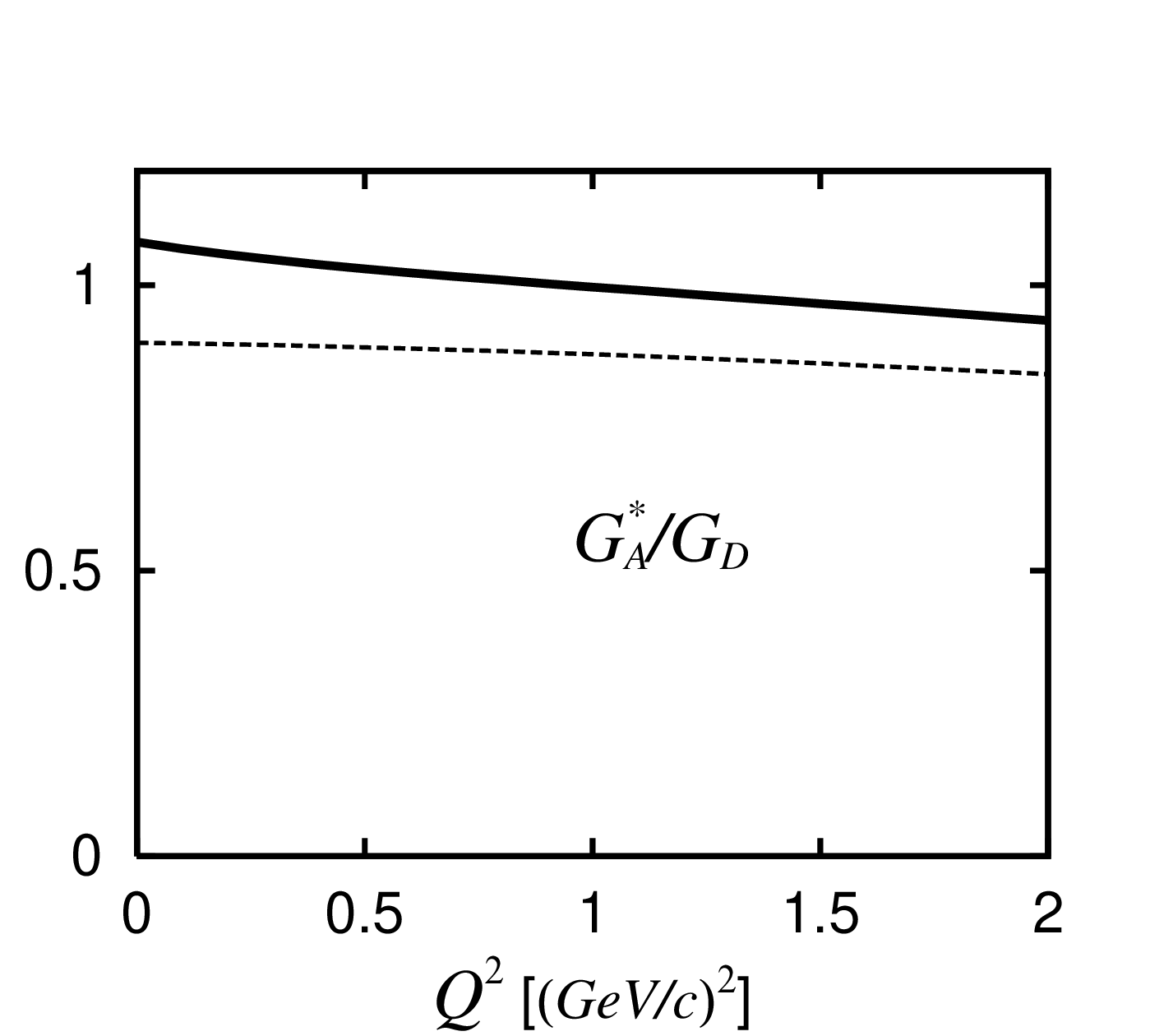}
\caption{$G_M$(left) and $G_A$(right).}
\end{figure}

\section{Neutrino-nucleus reaction}

We now apply the dynamical model explained in the previous section
to neutrino-nucleus reactions in the 
GeV region.
The charged-current neutrino-nucleus reaction cross section 
can be written as
\begin{eqnarray}
\frac{d\sigma}{dE'd\Omega'} &
 = & \frac{p_l'}{p_l}\frac{G_F^2 \cos^2\theta_c}{8\pi^2}L_{\mu\nu}W^{\mu\nu}
\end{eqnarray}
where $L^{\mu\nu}$ is
the lepton tensor.  The hadron
tensor  $W^{\mu\nu}$ is given as
\begin{eqnarray}
W^{\mu\nu} & = & \bar{\sum}_i \sum_f (2\pi)^3 \frac{E_T}{M_T} 
 \delta^4(q + P_i - P_f) \nonumber \\
 & &<\!f|J^\mu|i\!><\!f|J^\nu|i\!>^*,
\end{eqnarray}
where $P_{i,f}$ is the initial or final hadron momentum,
$q$ a momentum transfer.

As a first step of our neutrino-nucleus reaction study, 
we consider the following points:
 (1) Fermi-averaging of the elementary amplitude is
 taken into account by calculating the
reaction energy $W=\sqrt{(q+p_N)^2}$
of the elementary process from
the nucleon momentum $p_N$ inside nucleus and momentum transfer $q$.
(2) We also take into account Pauli effects 
on
the final nucleon after pion production. We require
that the momentum of the
final nucleon $|\vec{p}'|=|\vec{p}_N + \vec{q} -\vec{k}|$ 
be larger than the Fermi momentum.
This correction is calculated
for each pion momentum $\vec{k}$ and initial 
nucleon momentum $\vec{p}_N$.
Fig. 6 shows the cross section for the
$\nu_e + ^{12}C \rightarrow e^- + \pi + N + (A=11)$ reaction
divided by mass number
for $E_\nu=1$ GeV and for the two values of the lepton scattering
angle, 10$^o$ and 30$^o$.
The dash-dotted curve shows the average of the
cross sections for the proton and neutron targets,
$(\sigma_p + \sigma_n)/2$.
Fermi averaging is included in the dotted curve,
and furthermore the Pauli-effect
is included in the solid curve.
As discussed in Ref.~\cite{pascos2},
the Pauli effect is appreciable at low momentum transfers,
i.e. at forward lepton
angles. 
In evaluating the Pauli effect,
we take full account of the pion angular distribution 
as predicted by the dynamical model.

\begin{figure}
\includegraphics[width=3.7cm]{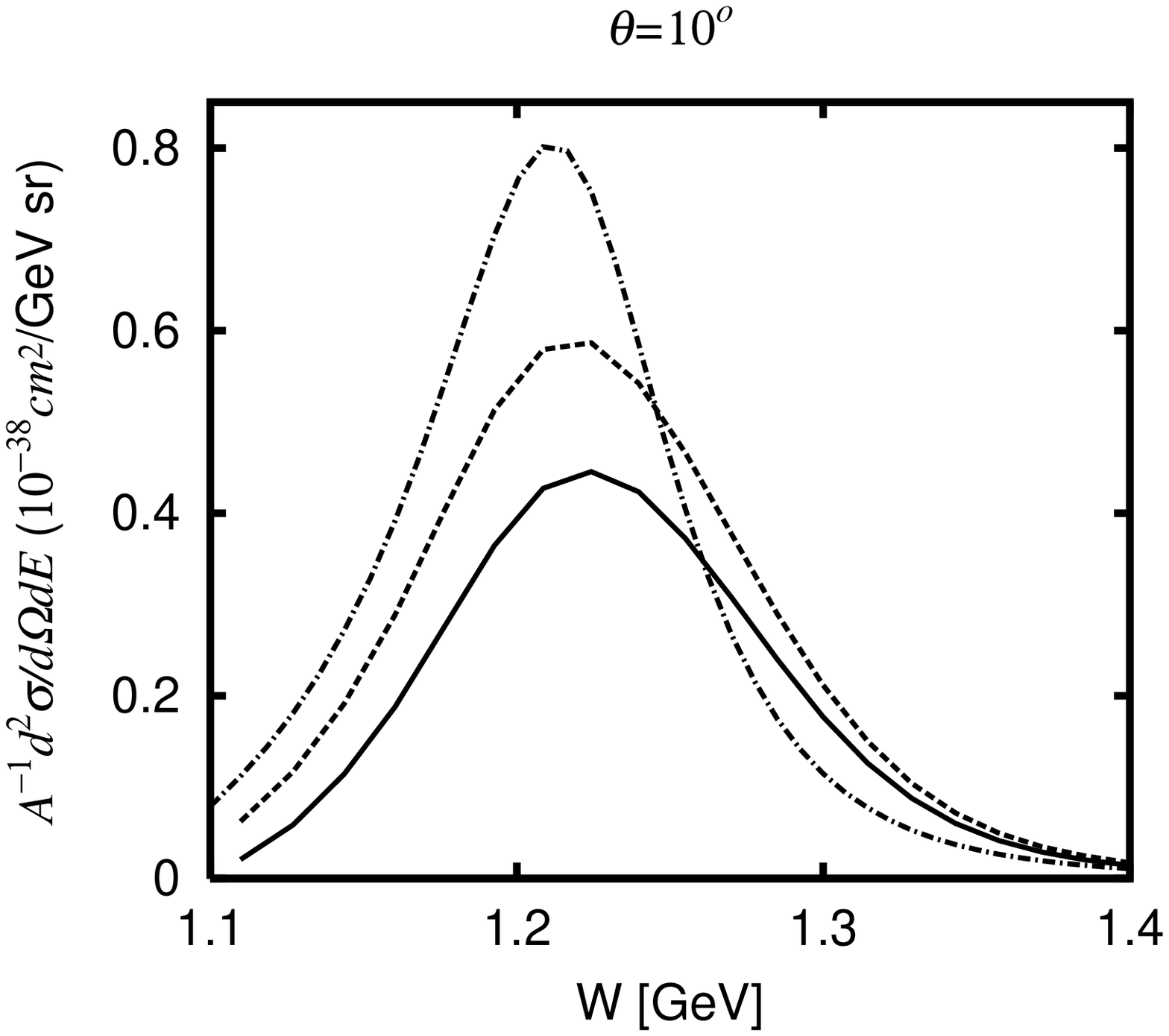}\hspace*{-0.1 cm}
\includegraphics[width=3.7cm]{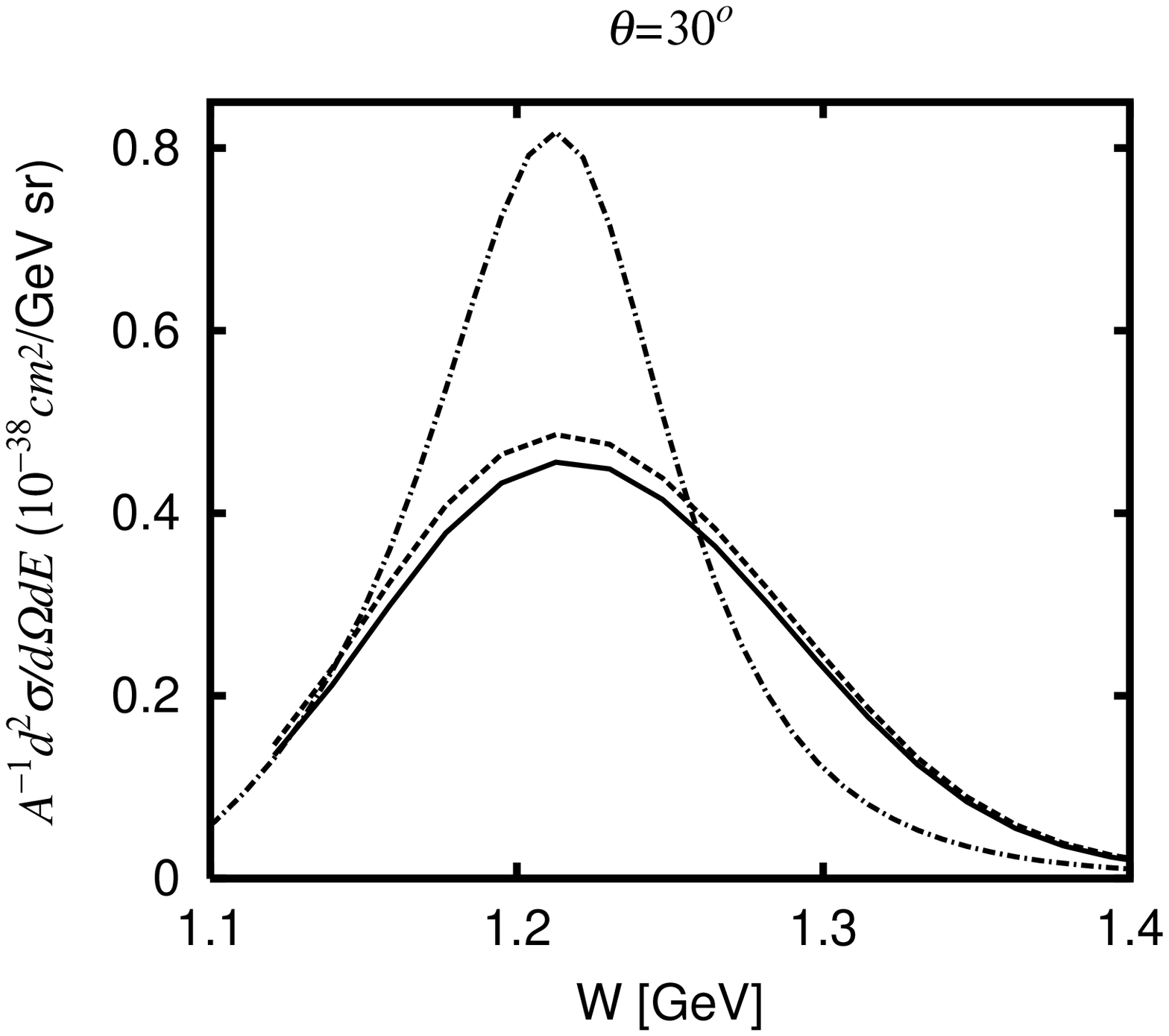}
\caption{Cross section of $\nu + ^{12}C 
\rightarrow e^- + \pi + X$.}
\end{figure}

To examine to what extent our treatment can describe the data,
we have studied inclusive electron scattering
on  $^{12}C$.
The quasi-elastic process is evaluated using 
the relevant 
formula in Ref.~\cite{moniz}.
The pion production process is calculated as described above.
 Here the nuclear correlation
is taken into account by using 
the structure function given in Ref.~\cite{benhar1}.
In Fig.\ 7, our results are
compared with the data
for $E_e=0.96$ and $1.1$ GeV  at $\theta=37.5^o$~\cite{sealock}.
The magnitude of the calculated
pion production cross section 
is comparable with the data,
while our result (solid line) tends to overestimate the quasi-free cross 
section.
Furthermore the 
`dip' region between the quasi-free and delta-excitation
peaks is not explained within the model.

Recently there have been detailed studies of nuclear effects
in inclusive neutrino-nucleus reactions~\cite{Benetal05,Nieetal05}.
Meanwhile, in the present  first attempt to apply the dynamical model
of Refs.~\cite{sl1,sl2,sl3,sl4}
to neutrino-nucleus reactions,
we have included nuclear-medium effects
only by incorporating the Fermi-averaging and Pauli effects
in the way described above
and by introducing the structure function. 
It is obviously important to investigate any possible 
nuclear effects beyond the level considered in our present work.

\begin{figure}[hbt]
\includegraphics[width=3.7cm]{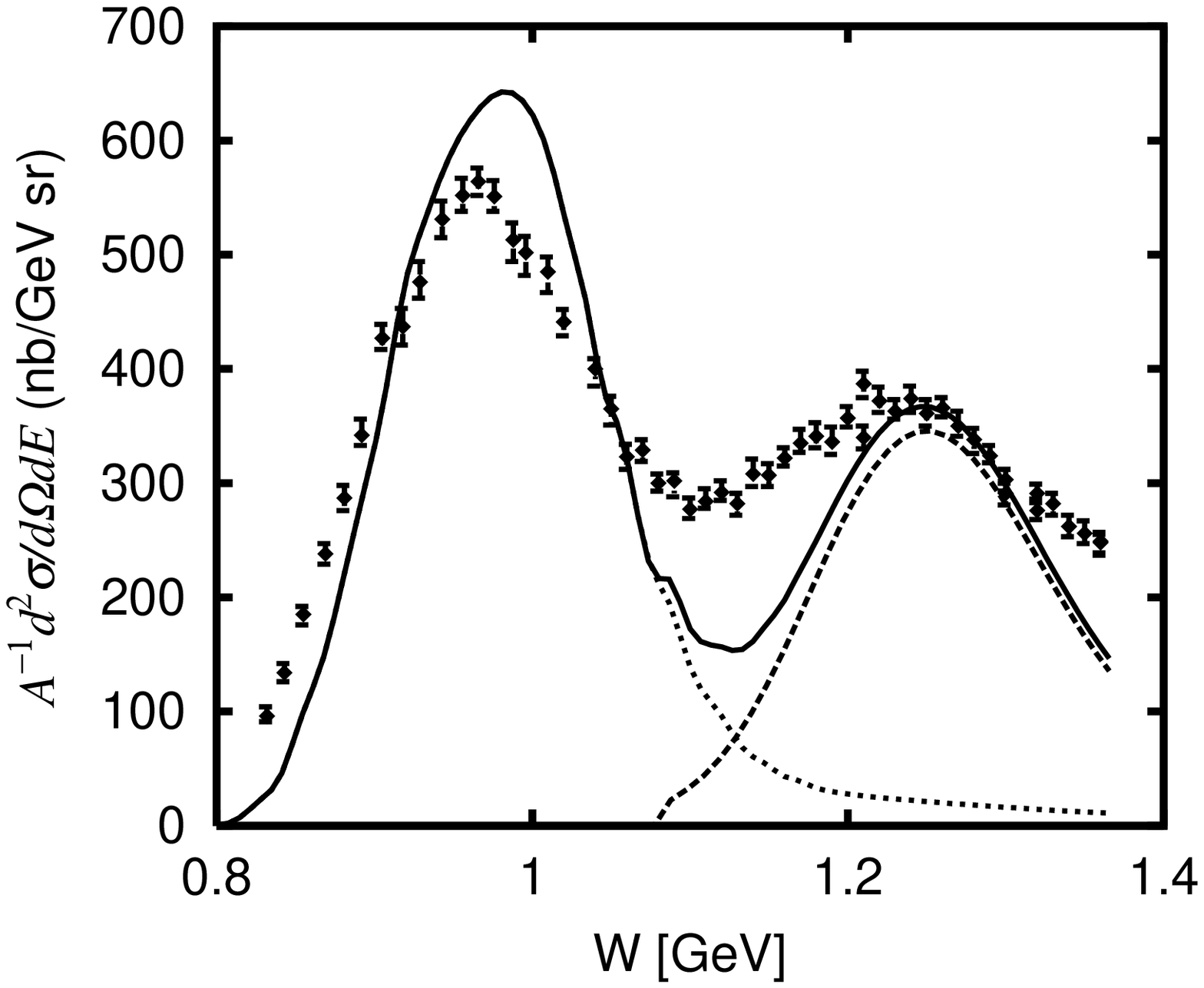}\hspace*{-0.1cm}
\includegraphics[width=3.7cm]{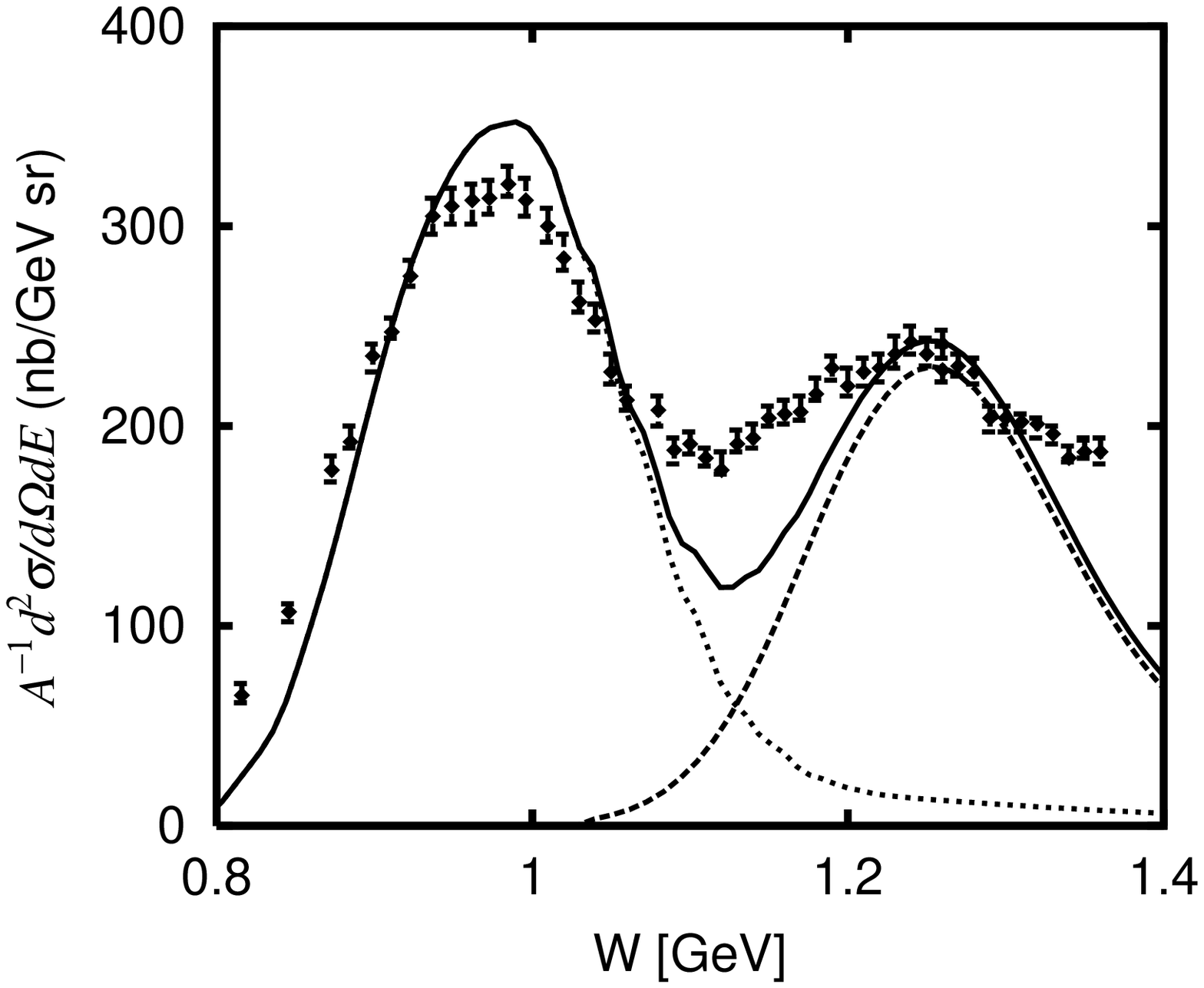}
\caption{Inclusive cross section of
 $^{12}C + e \rightarrow e' + X$
at $E_e=0.96$ GeV (left) and $E_e=1.1$ GeV (right).}
\end{figure}

\section{Summary}

We have developed a dynamical model of electroweak pion  
production reactions.
Most of the available
data of pion electroproduction and neutrino  reaction
on the nucleon in the $\Delta$ region  can be described reasonably well
by the model.
The $N\Delta$ transition form factors 
extracted from the data exhibit
a large contribution of the
pion cloud effects to the $N\Delta$ form factors,
and these effects explain why the simple quark models
fail to explain the magnitude and the $Q^2$ dependence 
of the empirical form factors.
We have studied neutrino-nucleus reactions
using the dynamical model that has been well tested
by the electron scattering data.
We have examined the effects of Fermi motion,
Pauli blocking and nuclear correlations
using the dynamical model. 
It is shown that these effects are important
and significantly improve 
the description
of the electron-nucleus scattering around 1GeV.
However further  
studies on the propagation of the $\Delta$-particle
in nuclei seem to be needed.

\section*{Acknowledgments}
 
This work is supported by the Japan Society for the Promotion of Science,
Grant-in-Aid for Scientific Research(c) 15540275,
by the U.S. National Science Foundation, Grant Nos. PHY-0140214
and PHY-0457014, and by the U.S. Department of Energy,
Nuclear Physics Division Contract No. W-31-109-ENG-38.

\end{document}